\input{aipcheck}
\documentclass[
    ,final            
  ]
  {aipproc}
\layoutstyle{6x9}
\begin{document}
\title{Dynamic Critical Behavior of Percolation Observables in 
       the 2d Ising Model}

\author{W. G. Wanzeller}{
  address={Instituto de F\'{\i}sica Te\'orica,
  Universidade Estadual Paulista\\
  Rua Pamplona 145,
  01405-900, S\~ao Paulo, SP, Brazil}
}
\author{G. Krein}{
  address={Instituto de F\'{\i}sica Te\'orica,
  Universidade Estadual Paulista\\
  Rua Pamplona 145, 
  01405-900, S\~ao Paulo, SP, Brazil}
}
\author{T. Mendes}{
  address={Instituto de F\'{\i}sica de S\~ao Carlos,
  Universidade de S\~ao Paulo \\
  C.P. 369, 13560-970, S\~ao Carlos, SP, Brazil }
}
\begin{abstract}
We present preliminary results of our numerical study of the critical 
dynamics of percolation observables for the two-dimensional Ising model.
We consider the (Monte-Carlo) short-time evolution of the system 
obtained with a local heat-bath method and with the global 
Swendsen-Wang algorithm. 
In both cases, we find qualitatively different dynamic behaviors for
the magnetization and $\Omega$, the order parameter of the percolation
transition. This may have implications for the recent attempts to
describe the dynamics of the QCD phase transition using cluster
observables.
\end{abstract}

\maketitle

\section{Introduction}
\vspace{-0.3 cm}
The study of the dynamic critical behavior of simple statistical
models might be of relevance for understanding
non-equilibrium effects in hot QCD, such as
the effects due to heating and
cooling of matter produced in heavy-ion collisions. 
The possible connection \cite{baym} between the deconfinement transition 
in QCD and the percolation phenomenon \cite{stauffer} has received 
renewed attention recently \cite{satz} and the dynamics of cluster
observables has been investigated using hysteresis methods \cite{hyst}.
As a preparation for the study of the continuous-spin $O(4)$ model,
whose magnetic transition is expected to be in the same universality
class of the chiral phase transition in two-flavor QCD, we consider 
here the short-time dynamics of the 2d Ising model and focus on
the dynamic critical behavior of percolation observables.

For many physical systems, a suitable definition of {\em cluster} 
provides a mapping of the physical phase transition into the geometric
problem of percolation, allowing a better understanding of how the
transition is induced in the system. 
For Ising and $O(N)$ spin models this mapping 
is well understood \cite{BJP,blanchard}, whereas in QCD it may be
harder to define, even in the pure-gauge case \cite{satz}.
Given a definition for a cluster on the lattice, the
order parameter in percolation theory is the 
{\em stress of the percolating cluster} $\Omega$, defined by
\begin{eqnarray}
\Omega \equiv \left \{
\begin{array}{ccl}
0  \,, & T >T_c  &\quad  \\
\Delta/V \,, & T \leq T_c & \quad 
\end{array} \right .
\end{eqnarray}
where $\Delta$ is the volume of the percolating cluster\footnote{Note 
that a percolating cluster is a set of spins 
connected from the first to the last row of the lattice.}
and {$T_c$} is the critical temperature.

\section{Short-time (Monte Carlo) dynamics}
Using renormalization-group theory, it can be shown \cite{janssen}
that the early time evolution of an order parameter (e.g.\ the magnetization
$M$) already displays universal critical behavior, given by
\begin{eqnarray}
M(t,\epsilon,m_0,L) = L^{-\beta/\nu}{\cal M}(tL^{-z},\epsilon
L^{1/\nu},m_0L^{x_0}) \,,
\end{eqnarray} 
where $\epsilon \equiv (T-T_c)/T_c \,$ and ${\cal M}$ is a universal 
function. We thus expect for $T=T_c$ a power-law 
behavior at early times $\,M(t)_{\epsilon \rightarrow 0} \sim m_0 t^\theta$, 
with $\theta=(x_0-\beta/\nu)/z\,$.
In principle, we would assume that the percolation order parameter 
{$\Omega$} should have a similar behavior.
The time evolution for the heat-bath and cluster algorithms is described 
below.

The heat-bath dynamics \cite{landau} consists in choosing the two possible
directions of each Ising spin according to the exact conditional probability
given by its nearest neighbors.
Each spin $S_i$ is chosen {\em UP} with probability
${p_i}$ , or {\em DOWN} with probability  ${1-p_i}$, where
\begin{eqnarray}
p_i = \frac{1}{1+\exp(-2\beta \sum_{j} S_j)}
\end{eqnarray} 
and the sum is over nearest neighbors of $S_i$.
After a certain number of iterations the spin configuration obeys the 
Boltzmann distribution. In the heat-bath method, since the updates are 
local, this transient time becomes considerably large at criticality.

The Swendsen-Wang (cluster) algorithm is obtained from
the Ising-model Hamiltonian by writing the partition function as
\begin{eqnarray}
{\cal Z}=\sum_{\{S\}}\sum_{\{n\}}\left\{
\prod_{\langle i,j\rangle}^{n_{ij}=1} p_{ij}\delta_{S_iS_j}\right \}
\left\{ \prod_{\langle i,j\rangle}^{n_{ij}=0}(1-p_{ij})   \right \}\,,
\end{eqnarray}
where $p_{ij}=1-\exp(-2J\beta)$ is the probability of having a
link between two nearest-neighbor sites of equal spin value. This
link is represented by ${n_{ij}}$ and determines the
clusters that will be associated with percolation at the critical 
temperature \cite{sokal}. The dynamics then consists of {\em global} 
moves in which the spins of a cluster are flipped together, 
with probability $1/2$.

\section{Numerical results}
In order to study the short-time dynamics we simulate at {$T=T_c$}
and force the system to have an initial magnetization {$m_0$}.
We let the system evolve in time and look for power-law behavior of the
order parameters $M$ and $\Omega$ as a function of the (Monte Carlo) time.
Each temporal sequence is generated from a different random seed,
i.e.\ each sequence has a different initial spin configuration.
The time history is then obtained from an average over all the
generated sequences.
We have performed Monte Carlo simulations with 50000 seeds and 5000 sweeps 
for three initial magnetizations ($m_0=0.02$, 0.03, 0.04),
for six lattice volumes, for both heat-bath and Swendsen-Wang
algorithms.

\begin{figure}[t]
\protect\vspace*{-3cm}
\includegraphics[height=0.4\hsize,angle=-0]{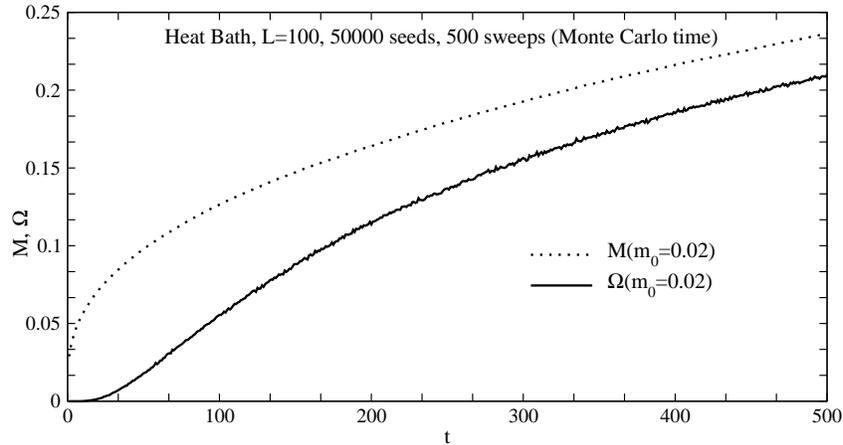}
\caption{
Plot of the early time evolution of the magnetic ($M$)
and percolation ($\Omega$) order parameters for the heat-bath case.}
\label{fig:omegaM}
\end{figure}

We obtain that a power-law fit works very well for $M$,
yielding the literature value \cite{zhang} for the exponent $\theta$.
However, as can be seen in Fig.\ \ref{fig:omegaM},
the percolation order parameter $\Omega$ does not show a power-law behavior,
being consistent with an exponential behavior of the type
$\exp(-\tau/t)$, where $\tau$ is essentially the relaxation time
to equilibrium. This seems to hold also for the case of the
Swendsen-Wang dynamics.
We thus see that although the equilibrium behaviors of $M$ and $\Omega$
are equivalent, the two order parameters show qualitatively different
dynamic critical behavior.

\begin{theacknowledgments}
This work was partially supported by CNPq, CAPES and FAPESP
(Brazilian financial agencies).
\end{theacknowledgments}

\bibliographystyle{aipproc}   
\bibliography{hadrons_proc}

\hyphenation{Post-Script Sprin-ger}
\begin{thebibliography}{10}
\expandafter\ifx\csname natexlab\endcsname\relax\def\natexlab#1{#1}\fi
\providecommand{\enquote}[1]{``#1''}
\expandafter\ifx\csname url\endcsname\relax
  \def\url#1{\texttt{#1}}\fi
\expandafter\ifx\csname urlprefix\endcsname\relax\def\urlprefix{URL }\fi

\bibitem[Baym(1979)]{baym}
Baym, G., \emph{Physica}, \textbf{A96}, 131 (1979).

\bibitem[Stauffer and Aharony(1992)]{stauffer}
Stauffer, D., and Aharony, A., \emph{Introduction to Percolation Theory},
  Taylor \& Francis, 1992.

\bibitem[Satz(2002)]{satz}
Satz, H., \emph{Comp.\ Phys.\ Comm.}, \textbf{147}, 46 (2002).

\bibitem[B.A.~Berg and Velytsky(2004)]{hyst}
Berg, B.A., Heller, U.M., Meyer-Ortmanns, H., and Velytsky, A.,
\emph{Phys.\ Rev.}, \textbf{D69}, 034501 (2004).

\bibitem[W.G.~Wanzeller and Mendes(2004)]{BJP}
Wanzeller, W.G., Cucchieri, A., Krein, G., and Mendes, T., \emph{Braz.\ J.\ Phys.},
  \textbf{34}, 247 (2004).

\bibitem[P.~Blanchard and Satz(2000)]{blanchard}
Blanchard, P., et al., \emph{J.\ Phys.}, \textbf{A33}, 8603 (2000).

\bibitem[H.K.~Janssen and Schmittmann(1989)]{janssen}
Janssen, H.K., Schaub, B., and Schmittmann, B.,
\emph{Z. Phys.}, \textbf{B73}, 539 (1989).

\bibitem[Landau and Binder(2000)]{landau}
Landau, D.~P., and Binder, K., \emph{Monte Carlo Simulations in Statistical
  Physics}, Cambridge, 2000.

\bibitem[Sokal(1996)]{sokal}
Sokal, A., \emph{Monte Carlo Methods in Statistical Mechanics: Foundations and
  New algorithms}, 1996,
  \urlprefix\url{http://citeseer.nj.nec.com/sokal96monte.html}.

\bibitem[J.-B. Zang. L.~Wang and Ji(1999)]{zhang}
Zang, J.-B., et al., \emph{Phys. Lett.}, \textbf{A262}, 226 (1999).

\end{thebibliography}
\IfFileExists{\jobname.bbl}{}
 {\typeout{}
  \typeout{******************************************}
  \typeout{** Please run "bibtex \jobname" to optain}
  \typeout{** the bibliography and then re-run LaTeX}
  \typeout{** twice to fix the references!}
  \typeout{******************************************}
  \typeout{}
 }

\end{document}